\documentclass[preprint, sort&compress]{elsarticle}
\usepackage{amsmath,amssymb,braket,multirow,url}
\usepackage{graphicx}
\usepackage{graphics}
\usepackage{subfigure}
\usepackage{txfonts}
\newcommand{\Ax}[1]{A_{#1}}
\newcommand{\Axd}[1]{A_{#1}^{\dagger}}
\newcommand{\Bx}[1]{B_{#1}}
\newcommand{\Bxd}[1]{B_{#1}^{\dagger}}
\newcommand{\bx}[1]{b_{#1}}
\newcommand{\bxd}[1]{b_{#1}^{\dagger}}
\newcommand{\Cx}[1]{C_{#1}}
\newcommand{\Cxd}[1]{C_{#1}^{\dagger}}
\newcommand{\cx}[1]{c_{#1}}
\newcommand{\cxd}[1]{c_{#1}^{\dagger}}
\newcommand{\Nbar}{\overline{N}}
\newcommand{\nbar}{\overline{n}}
\begin{document}
\title{Understanding finite size effects in quasi-long-range orders for exactly
solvable chain models}
\author{Sisi Tan}
\ead{s080063@e.ntu.edu.sg}
\author{Siew Ann Cheong}
\ead{cheongsa@ntu.edu.sg}
\address{Division of Physics and Applied Physics, School of Physical and
Mathematical Sciences, Nanyang Technological University, 21 Nanyang Link,
Singapore 637371, Republic of Singapore}
\date{\today}
\begin{abstract}
In this paper, we investigate how much of the numerical artefacts introduced by
finite system size and choice of boundary conditions can be removed by finite
size scaling, for strongly-correlated systems with quasi-long-range order.
Starting from the exact ground-state wave functions of hardcore bosons and
spinless fermions with infinite nearest-neighbor repulsion on finite periodic
chains and finite open chains, we compute the two-point, density-density, and pair-pair correlation functions, and fit these to various asymptotic power laws.  Comparing the
finite-periodic-chain and finite-open-chain correlations with their infinite-chain counterparts, we find reasonable agreement among them for the power-law amplitudes and exponents, but poor agreement for the phase shifts.  More importantly, for chain lengths on the order of 100, we find our finite-open-chain calculation overestimates some infinite-chain exponents (as did a recent density-matrix renormalization-group (DMRG) calculation on finite smooth chains), whereas our finite-periodic-chain calculation underestimates these exponents.  We attribute this systematic difference to the different choice of boundary conditions. Eventually, both finite-chain exponents approach the infinite-chain limit: by a chain length of 1000 for periodic chains, and $> 2000$ for open chains. There is, however, a misleading apparent finite size scaling convergence at shorter chain lengths, for both our finite-chain exponents, as well as the finite-smooth-chain exponents. Implications of this observation are discussed.
\end{abstract}
\maketitle
\section{Introduction}
Long-range and quasi-long-range orders in strongly-correlated systems are the
object of numerous theoretical \cite{Daniel1991PA, Tian2002PRB66e224408} and
numerical \cite{Hirsch1989PRB40e4769,Minoru1989PRB40e2494, Millis2000PRB61e12496} studies.  While it is fairly straightforward to
ascertain the presence or absence of true long-range order
\cite{Xavier2007PRB76e014422, Hohenberg1976PR158, mermin1966PRL17,
karl1992PRB45e7229, Hirsch1989PRL62e591}, it is much harder to identify the
dominant correlations from amongst several competing quasi-long-range orders in
the quantum-mechanical ground-state wave function.  Much of this difficulty
lies with the fact that few models that describe strongly-correlated systems
can be solved analytically, and we have to resort to approximate or numerical
methods to solve for the ground-state wave function.  In particular, numerical
solutions can only be obtained for a finite system, when our real interest is
in the dominant quasi-long-range order of the infinite thermodynamic system.
An obvious solution to this finite size problem would be to harness as much
computational power as we can, to simulate extremely large systems
\cite{Liebsch2009PRB80e165126, Hager2003jcp194, Yuki2009Jcp130e234114}.  Even
so, our state-of-the-art simulations are still many orders of magnitude smaller
than systems probed experimentally.  Alternatively, we can simulate systems of
different sizes, and thereafter perform finite size scaling
\cite{fisch2009PRB79e214429, Lee1991PRB43e1268, Chayes1986PRL57}, to
extrapolate the results to infinite system size.

In finite size scaling, the computational condensed matter community is
essentially guided by general results from statistical mechanics
\cite{Hofstetter1993EL21e993, Marcelo1988PRB38e9172,
Li1996PRE53e2940, Toral2007CCP2e177}, as well as exact solutions
\cite{Bernu1992PRL69e2590, Bethe1931ZP71e1312, CN1967PRL19e1312}. However,
finite size scaling is still very much an art, with no rigorous theorem proving
that it will always work. Besides the finite number of sites, our choice of
boundary conditions might also affect the convergence properties of finite size
scaling, and hence the reliability of our numerical solutions. In two recent
studies, various quasi-long-range correlation functions were calculated
using exact diagonalization \cite{Cheong2009PRB79e212402} and DMRG
\cite{Munder2009NJP0910}.  In particular, in
Ref.~\cite{Munder2009NJP0910}, the power-law exponents obtained after
finite size scaling do not agree with those derived from exact solutions in Ref.
\cite{Cheong2009PRB80e165124}.

In this paper, we investigate the relative importance of finite system size and
choice of boundary conditions, by comparing the power-law exponents of various
correlation functions in a finite periodic and open chain of hardcore bosons or spinless
fermions with infinite nearest-neighbor repulsion, against those of the infinite
chain,\cite{Cheong2009PRB80e165124} as well as exponents of finite smooth chain obtained from the DMRG
calculation \cite{Munder2009NJP0910}.  In Sec.~\ref{sect:methods}, we will
describe how the ground state of finite periodic/open chain of particles with
infinite nearest-neighbor repulsion (excluded chain) can be mapped to the ground
state of finite periodic/open chain of particles without nearest-neighbor repulsion
(included chain).  Based on this correspondence between ground states, we then
explain how correlation functions in the excluded chain can be written in terms
of corresponding correlation functions in the included chain, through an
intervening-particle expansion.  Thereafter, we present in
Sec.~\ref{sect:results} our results for the two-point, density-density, and
pair-pair correlation functions for finite excluded chains of hardcore bosons
and spinless fermions.  We fit these to different asymptotic combinations of
simple and oscillatory power laws, to find generally good agreement between the
amplitudes and exponents, and generally poor agreement between the phase shifts
for finite and infinite chains.  More importantly, we find for finite chains of the same intermediate lengths, reasonable agreement between our open-chain exponents and the finite-smooth-chain exponents in DMRG study. Both show strong systematic differences with the finite-periodic-chain exponents. By repeating these exponent calculations for very long chains, we established that these differences at intermediate chain lengths to be due to the different choice of boundary conditions. These findings are then summarized in Sec.~\ref{sect:conclusions}.

\section{Models and Methods}
\label{sect:methods}
This section is organized into two subsections. In Sec.~\ref{sect:chainmodels},
we define our chain models of hardcore bosons and spinless fermions with
infinite nearest-neighbor repulsion, and show how the ground states of these
finite periodic/open excluded chains can be mapped to the ground states of finite
periodic/open included chains of hardcore bosons and spinless fermions without
nearest-neighbor repulsion through a right-exclusion map.  Then in
Sec.~\ref{sect:interpartexp}, we describe the intervening-particle expansion
method, which allows us to write the ground-state expectation of a given
excluded-chain observable, as a sum over conditional ground-state expectations
of included chain observables.

\subsection{Chain models}
\label{sect:chainmodels}
The excluded chain models for hardcore bosons and spinless fermions are given by
the Hamiltonians
\begin{equation}
\begin{aligned}
H^{(e,b)} &= -t \sum_j \left(\Bxd{j}\Bx{j+1} + \Bxd{j+1}\Bx{j}\right) +
V \sum_j N_j N_{j+1} + {} \\
&\quad\ U \sum_j N_j (N_j - 1), \\
H^{(e,f)} &= -t \sum_j \left(\Cxd{j}\Cx{j+1} + \Cxd{j+1}\Cx{j}\right) +
V \sum_j N_j N_{j+1},
\end{aligned}
\end{equation}
respectively, where $t$ is the hopping matrix element, $V$ is the
nearest-neighbor repulsion, and $U$ is the on-site repulsion for hardcore
bosons.  Exact solutions for these models can be obtained in the limit $U \to
\infty$ and $V \to \infty$, such that each site can only be singly occupied, and
no adjacent sites can be simultaneously occupied.

\begin{figure}[htbp]
\centering
\includegraphics[scale=0.25]{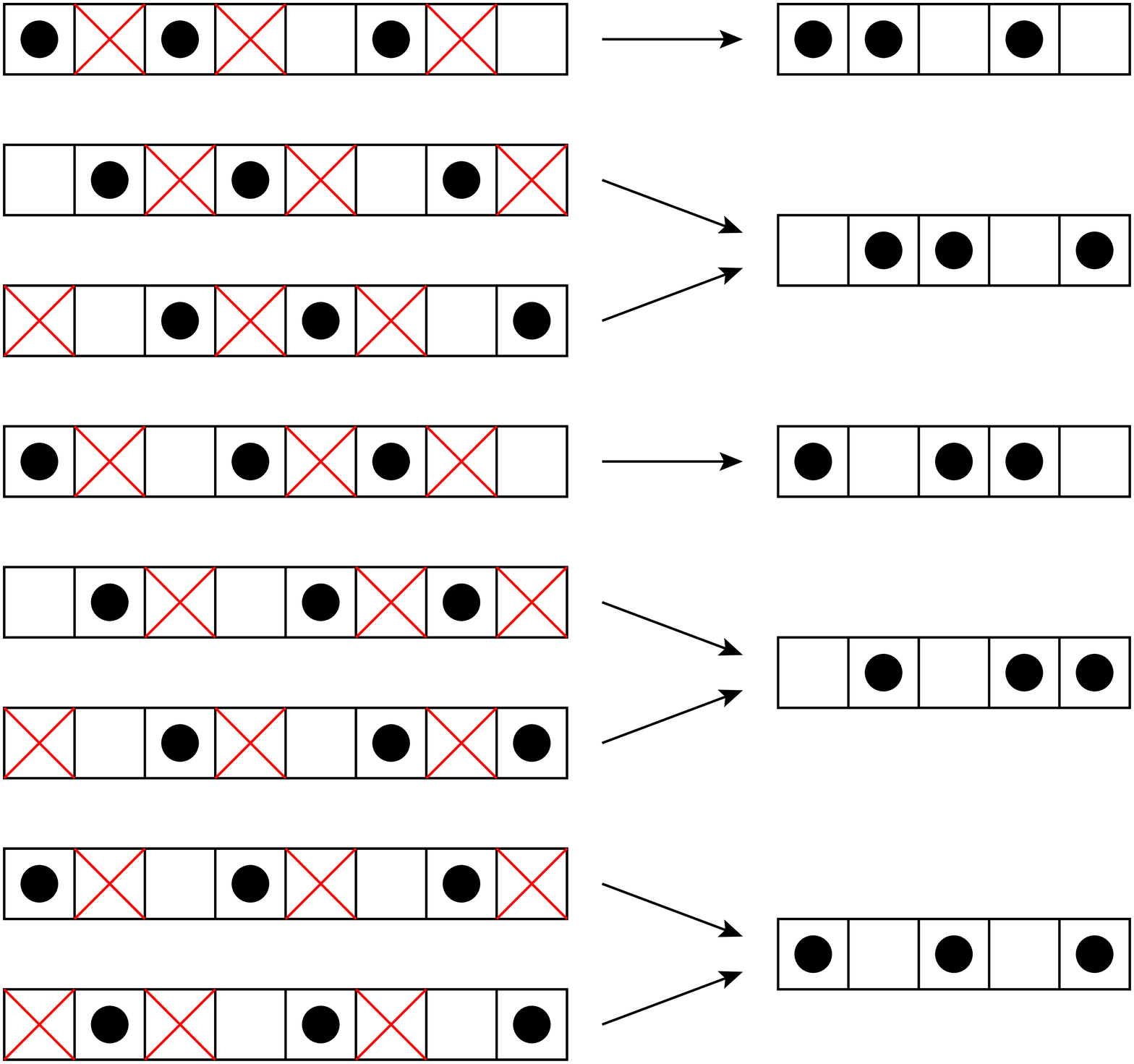}
\caption{Schematic diagram illustrating the many-to-one mapping from
$P$-particle configurations of an excluded periodic chain of length $L$ to
$P$-particle configurations of an included periodic chain of length $L' = L -
P$, by deleting a single empty site on the right of a particle.}
\label{fig:rightexclusionmap}
\end{figure}

The excluded chains can be mapped to included chains of hardcore bosons and
spinless fermions, with Hamiltonians
\begin{equation}
\begin{aligned}
H^{(i,b)} &= -t \sum_j \left(\bxd{j}\bx{j+1} + \bxd{j+1}\bx{j}\right) + U \sum_j n_j (n_j - 1), \\
H^{(i,f)} &= -t \sum_j \left(\cxd{j}\cx{j+1} + \cxd{j+1}\cx{j}\right).
\end{aligned}
\end{equation}
respectively, using a right-exclusion map first used by Fendley to map a
supersymmetric chain of spinless fermions to the $XXZ$ chain \cite{fendley03}.
As shown in Fig.~\ref{fig:rightexclusionmap}, the right-exclusion map deletes an
empty site to the right of each occupied site, to map an excluded configuration
for a chain of length $L$ to an included configuration for a chain of length $L
- P$. For finite open chains, this mapping is one-to-one, as shown in Ref.~\cite{Cheong2009PRB80e165124}.  For finite periodic chains, the right-exclusion map is not one-to-one, as illustrated in Fig. 1. Nevertheless, it is possible to construct a one-to-one correspondence between the Bloch states of the excluded chain and Bloch states of the included chain.  Details for this construction can be found in Ref.~\cite{Cheong2006thesis}.

For the rest of this paper, we will use $L$ to denote the length of the periodic/open
excluded chain, whose sites are indexed by $j = 1, \dots, L$.  We will also use
$P$ to denote the total number of particles on the chain, and $\Ax{j} = \Bx{j},
\Cx{j}$ as a common notation for hardcore boson and spinless fermion operators.

\subsection{Intervening-particle expansion}
\label{sect:interpartexp}
Because of the one-to-one correspondence between Bloch states, the Bloch-state
amplitudes in the excluded chain ground state are identical to those in the
included chain ground state.  Hardcore bosons or spinless fermions, the
included-chain ground state can be ultimately written in terms of a one-dimensional Fermi sea with discrete wave numbers.  Hence the ground-state expectations of all
observables on the excluded chain can ultimately be expressed in terms of such
Fermi-sea expectations.  In this subsection, we briefly describe how the
correlations $\braket{O_1 O_2}$ of two separated local operators of the
excluded chain can be calculated using the method of \emph{intervening-particle
expansion}.

First, for every observable $O$ on the excluded chain, we note that it is
possible to define a \emph{corresponding observable} $O'$ on the included chain,
such that
\begin{equation}\label{eqn:OOp}
\frac{1}{\Nbar} \braket{ J^{\alpha} | O | J^{\beta} } =
\frac{1}{\nbar} \braket{ j^{\alpha} | O' | j^{\beta} },
\end{equation}
for all excluded configurations $\ket{J^{\alpha}}$ and $\ket{J^{\beta}}$ that
are mapped to the included configurations $\ket{j^{\alpha}}$ and
$\ket{j^{\beta}}$ by the right-exclusion map.  Eq.~\eqref{eqn:OOp} then tells us
that
\begin{equation}
\frac{1}{\Nbar} \braket{O} = \frac{1}{\nbar} \braket{O'}
\end{equation}
between the ground-state expectations of $O$ and $O'$.

Next, we expand the ground-state expectation
\begin{equation}
\braket{O_j O_{j+r}} = \sum_{\{p\}} \braket{O_j O_p O_{j+r}}
\end{equation}
as a sum over conditional ground-state expectations, where $O_p$ is a product of
$p$ particle-occupation number operators $N_j$ and $r - p$ hole-occupation
number operators $(1 - N_j)$.  We call this sum over all possible ways to insert
particles between $O_j$ and $O_{j+r}$ the \emph{intervening-particle
expansion}.  Finally, we rewrite each conditional ground-state expectation
$\braket{O_j O_p O_{j+r}}$ on the excluded chain in terms of their
corresponding conditional ground-state expectation $\braket{O'_j O'_p
O'_{j+r-p}}$ on the included chain.  In terms of these conditional
expectations, which can in turn be written in terms of expectations of the
one-dimensional Fermi sea, the intervening-particle expansion becomes
\begin{equation}
\braket{O_j O_{j+r}} = \frac{\Nbar}{\nbar} \sum_{\{p\}}
\braket{O'_j O'_p O'_{j+r-p}},
\end{equation}
which can always be evaluated numerically for separations $r$ that are not too
large.

\section{Results and discussions}
\label{sect:results}
In Ref.~\cite{Cheong2009PRB80e165124}, three correlation functions: (i)
the two-point function $\braket{\Axd{j}\Ax{j+r}}$; (ii) the density-density
correlation function $\braket{N_j N_{j+r}}$; and (iii) the pair-pair correlation
function $\braket{\Axd{j}\Axd{j+2}\Ax{j+r}\Ax{j+r+2}}$ were systematically
examined for infinite chains of hardcore bosons and spinless fermions with
infinite nearest-neighbor repulsion, and also in three limiting cases for an
infinite ladder.  In this section, we calculate these three correlation
functions for the finite periodic/open chain, for comparison against those of the
infinite chain \cite{Cheong2009PRB80e165124}, as well as against those of finite smooth chain computed
in the recent DMRG study \cite{Munder2009NJP0910}. For the finite open chain, we analyze the numerical correlation functions only for $\bar{N} \leq 0.25$.  We believe the results are not reliable for $\bar{N} > 0.25$, because of phase separation in the finite open chain ground states (the symptom of which can be seen in Fig.~\ref{fig:cdwpofp}).

To understand the relative importance of finite chain length and choice of
boundary conditions, we fit all the slowly-decaying correlation functions to
the asymptotic form
\begin{equation}
A_0 + A_1\, r^{-A_2} + A_3\, r^{-A_4}\, \cos(k r + \phi),
\end{equation}
where $A_0$ is a constant term, $A_1$ and $A_2$ are the amplitude and exponent
of a simple power law, while $A_3$ and $A_4$ are the amplitude and exponent of
an oscillatory power law with wave number $k$ and phase shift $\phi$.  We then
check how these finite-chain exponents depend on chain length and boundary conditions, and thus the effectiveness of finite size scaling.

\subsection{Two-point functions}
\begin{figure}
\centering
\subfigure{\label{fig:flb}\includegraphics[width=0.45\textwidth]{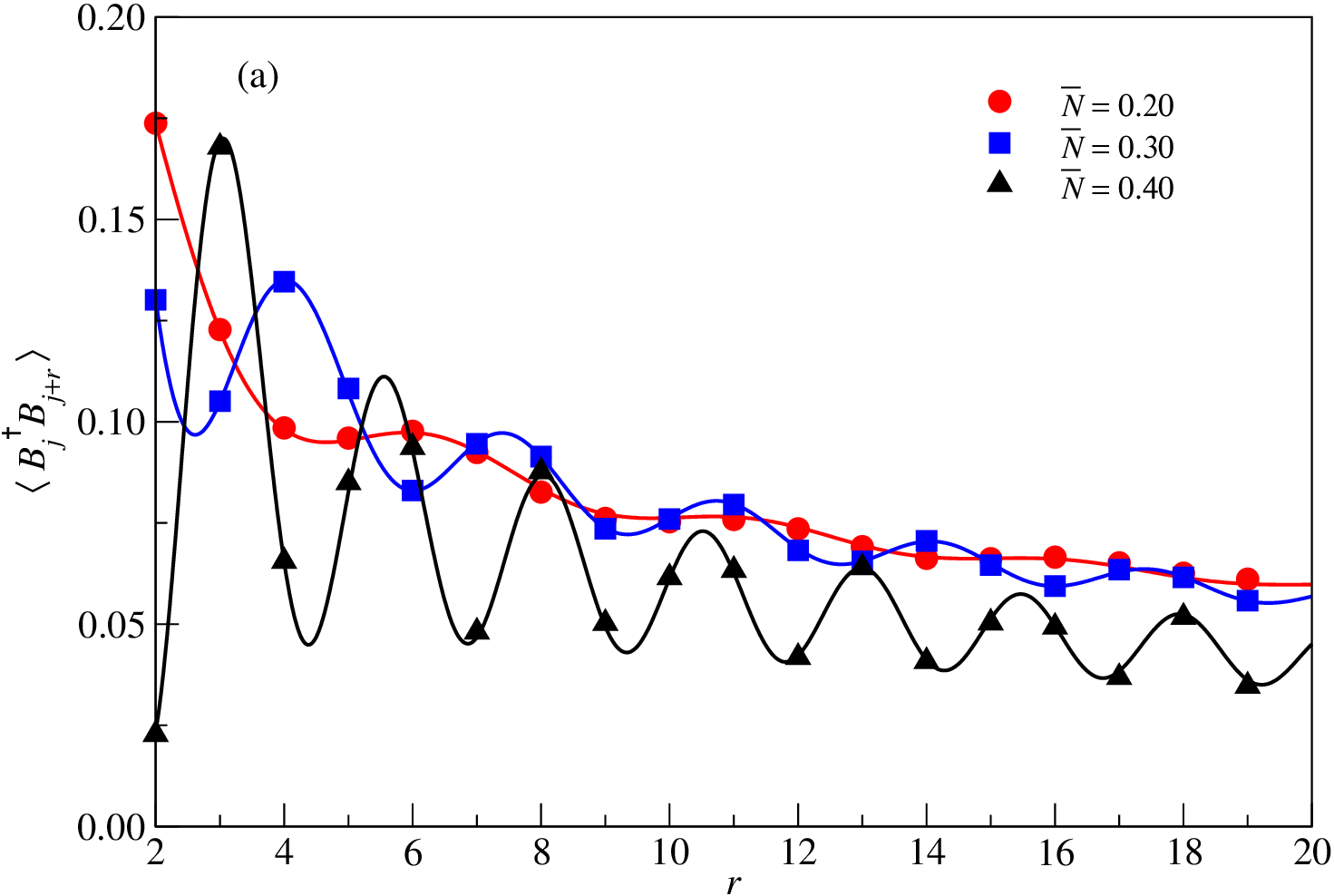}}
\subfigure{\label{fig:flbol3}\includegraphics[width=0.45\textwidth]{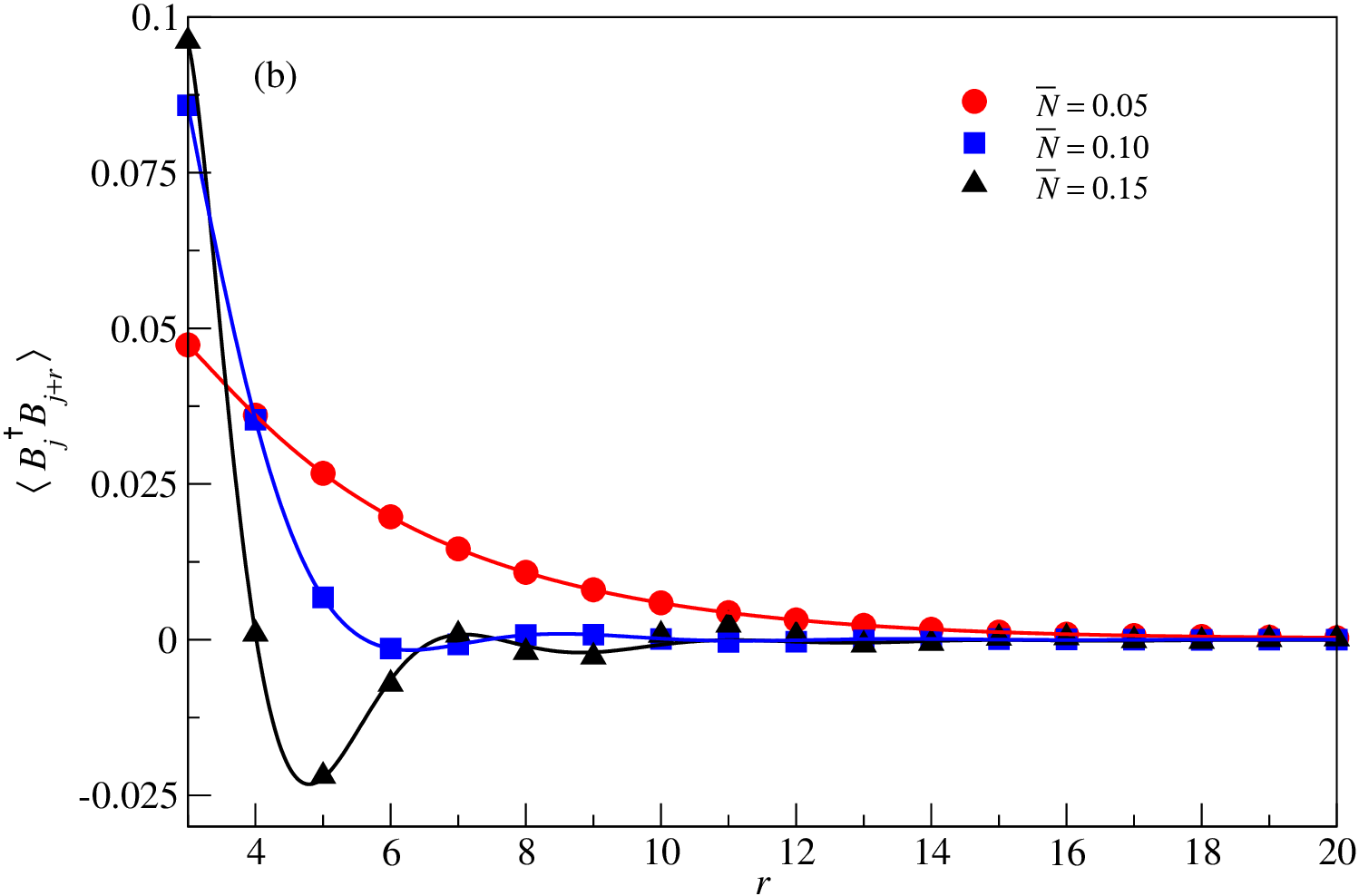}}
\caption{Two-point function $\braket{\Bxd{j}\Bx{j+r}}$ of hardcore bosons with
infinite nearest-neighbor repulsion on (a) a periodic and (b) an open chain of length $L = 100$ as a
function of the separation $r$.  In this figure, red circles, blue aquares, and
black triangles are the exact correlations at densities (a) $\Nbar = 0.20, 0.30, 0.40$ and (b) $\Nbar=0.05, 0.10,0.15$,
respectively. The red, blue and black curves are the best nonlinear fits of the
form $A_1\, r^{-A_2} + A_3\, r^{-A_4}\, \cos(k r + \phi)$ to the exact
correlations at these densities.}
\label{fig:flbpo}
\end{figure}

The hardcore boson two-point function $\braket{\Bxd{j}\Bx{j+r}}$ obtained using
the intervening-particle expansion appears to be an oscillatory power law
sitting on top of a simple power law.  As shown in Fig.~\ref{fig:flbpo}, we fit
the hardcore boson two-point function to the asymptotic form
$\braket{\Bxd{j}\Bx{j+r}} = A_1\, r^{-A_2} + A_3\, r^{-A_4}\, \cos(k r +
\phi)$.  As expected, the fitted wave number is $k = \pi\Nbar$.

\begin{figure}
\centering
\includegraphics[scale=0.325,clip=true]{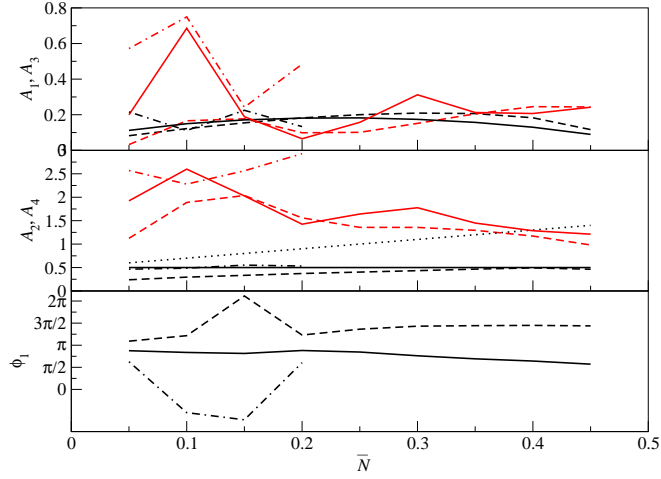}
\caption{The fitted (top) amplitudes $A_1$ (black) and $A_3$ (red), (middle) exponents $A_2$ (black) and $A_4$ (red), and (bottom) phase shift $\phi$ of the simple power law and oscillatory
power law as a function of density $\Nbar$, for the two-point function of a $L =
100$ periodic(dashed lines) and open(dot-dashed lines) chain of hardcore bosons with infinite nearest-neighbor repulsion. The corresponding parameters of the infinite chain from
Ref.~\cite{Cheong2009PRB80e165124} are shown as solid lines, whereas finite-smooth-chain results from Ref.~\cite{Munder2009NJP0910} are shown as dotted lines.}
\label{fig:flbpofp}
\end{figure}

For the infinite chain, the leading exponent (that of the simple power law) was
found to be very close to $A_2 = \frac{1}{2}$, as predicted by Efetov and
Larkin for an included chain of hardcore bosons \cite{Efetov1976SP42e390}.  For
our finite  periodic chain, this leading exponent was found to approach $\frac{1}{2}$
with increasing density from below, as shown in Fig.~\ref{fig:flbpofp}.  In
comparison, the finite-smooth-chain exponent $A_2 = 2\Nbar + \frac{1}{2}$ obtained in
Ref.~\cite{Munder2009NJP0910} approaches $\frac{1}{2}$ with decreasing
density from above. The finite-open-chain exponent $A_2$ agrees very closely with the infinite-chain results. Otherwise, we find reasonable agreement between the
amplitudes and exponents of the finite and infinite chains, but poor agreement
between their phase shifts.

Similarly, for the spinless fermion two-point function, which is fitted to a
single oscillatory power law $\braket{\Cxd{j}\Cx{j+r}} = A_3\, r^{-A_4}\, \cos(k
r + \phi)$ (see Fig.~\ref{fig:flfpo}), we find reasonable agreement between the
amplitudes and exponents of the finite and infinite chains, but poor agreement
between the phase shifts of the finite and infinite chains (see
Fig.~\ref{fig:flfpofp}).  As expected, the fitted wave number is $k = \pi\Nbar$.
For noninteracting spinless fermions, the exponent of the oscillatory power law
is $A_4 = 1$.  We see from Fig.~\ref{fig:flfpofp} that for both finite and
infinite chains, $A_4$ starts close to 1 at $\Nbar = 0$, and decreases with
increasing density.  In particular, the finite-chain exponent is smaller than
the infinite-chain exponent, telling us that two-point correlations are stronger
on finite chains.

\begin{figure}
\centering
\subfigure{\label{fig:flf}\includegraphics[width=0.45\textwidth]{flf3}}
\subfigure{\label{fig:flfol3}\includegraphics[width=0.48\textwidth]{flfol3}}
\caption{Two-point function $\braket{\Cxd{j}\Cx{j+r}}$ of spinless fermions with
infinite nearest-neighbor repulsion on (a) a periodic and (b) an open chain of length $L = 100$ as a
function of the separation $r$.  In this figure, red circles, blue squares, and
black triangles are the exact correlations at densities (a) $\Nbar = 0.2, 0.3, 0.4$ (b) $\Nbar=0.05, 0.10,0.15$, respectively. The red, blue and black curves are the best nonlinear fits of the form $A_3\, r^{-A_4}\, \cos(k r + \phi)$ to the exact correlations at these densities.}
\label{fig:flfpo}
\end{figure}

\begin{figure}
\centering
\includegraphics[scale=0.325,clip=true]{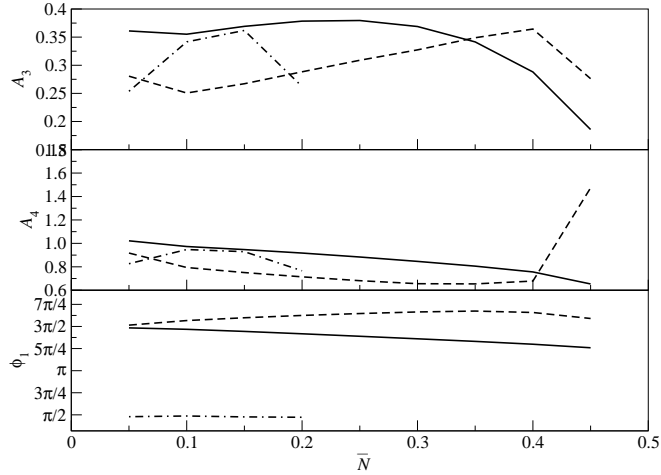}
\caption{The fitted amplitude $A_3$ (top), exponent $A_4$ (middle) and phase
shift $\phi$ (bottom) of the oscillatory power-law fit to the two-point function
of a $L = 100$ periodic(dashed lines) and open(dot-dashed lines) chain of spinless fermions with infinite
nearest-neighbor repulsion, as a function of density $\Nbar$.  The corresponding
parameters of the infinite chain from Ref.~\cite{Cheong2009PRB80e165124} are shown as solid lines.}
\label{fig:flfpofp}
\end{figure}

\subsection{Density-density correlations}
\begin{figure}
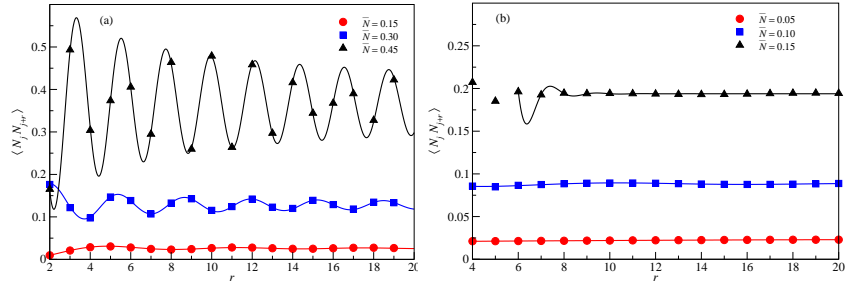

\centering
\subfigure{\label{fig:cdw3}\includegraphics[width=0.45\textwidth]{cdw1003}}
\subfigure{\label{fig:cdwol3}\includegraphics[width=0.458\textwidth]{cdwol3}}
\caption{Density-density correlation function $\braket{N_j N_{j+r}}$ of
particles with infinite nearest-neighbor repulsion on a periodic/open chain of length
$L = 100$ as a function of the separation $r$. In this figure, red circles, blue
squares, and black triangles are the exact correlations at densities (a) $\Nbar =
0.15, 0.30, 0.45$, and (b) $\Nbar=0.05, 0.10,0.15$, respectively. The red, blue and black curves are the best nonlinear fits of the form $A_0 + A_3\, r^{-A_4} \cos(kr + \phi)$ to the exact correlations at these densities.}
\label{fig:cdwpo}
\end{figure}

\begin{figure}
\centering
\includegraphics[scale=0.325,clip=true]{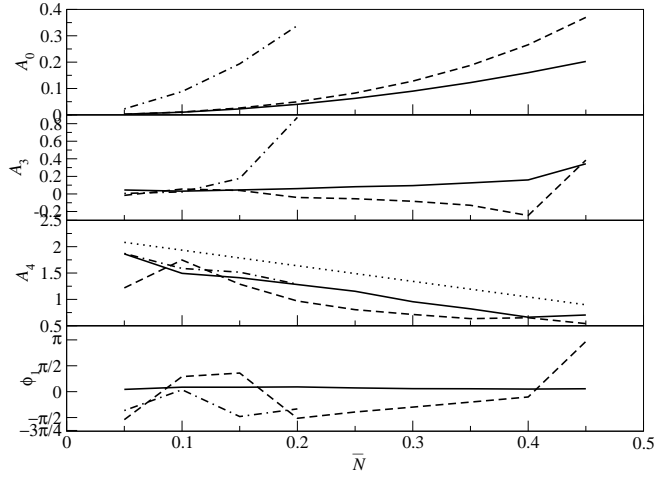}
\caption{The fitted constant $A_0$ (top), and amplitude $A_3$ (second from top),
exponent $A_4$ (second from bottom), and phase shift $\phi$ of the oscillatory
power law as a function of density $\Nbar$, for the density-density correlation
function of a $L = 100$ periodic(dashed lines) and open(dot-dashed lines) chain of particles with infinite
nearest-neighbor repulsion.  The corresponding parameters of the infinite chain
from Ref.~\cite{Cheong2009PRB80e165124} are shown as solid lines ($A_4 =
\frac{1}{2} + \frac{5}{2}(\frac{1}{2} - \Nbar)$ for the infinite chain), while
the finite-smooth-chain exponent $A_4 = -2.96\, \Nbar + 2.23$ is shown as the dotted line.}
\label{fig:cdwpofp}
\end{figure}

The density-density correlation function, which is identical for hardcore bosons
and spinless fermions on a chain, can be fitted very well to the asymptotic form
$\braket{N_j N_j+r} = A_0 + A_3\, r^{-A_4}\, \cos(k r + \phi)$ (see
Fig.~\ref{fig:cdwpo}).  Again, the fitted wave number $k = 2\pi\Nbar$ agrees with
our expectations.  Also, as with the two-point functions, we see in
Fig.~\ref{fig:cdwpofp} that there is reasonable agreement between the finite and
infinite chains for the fitted amplitudes and exponents, but poor agreement for
the phase shifts.  For both finite chains, we find $A_0 > \braket{N_j}\braket{N_j+r} = \Nbar^2$, the limit attained in the infinite excluded chain.  This excess density-density correlation compared to the
infinite chain persists at $L = 200$ and $L = 1000$.  Also shown in Fig.~\ref{fig:cdwpofp} are
the oscillatory power-law exponent $A_4$ for the finite periodic/open chain, the
infinite chain, and the finite smooth chain.

In Fig.~\ref{fig:cdwpofp}(a), while we see the exponents in all three cases decaying with increasing $\Nbar$, we again find the finite-periodic-chain exponent (dashed line) to be generally smaller
than the infinite-chain exponent (solid line), the finite-open-chain exponent to be slightly larger than the infinite-chain exponent, and the finite-smooth-chain exponent (dotted line) to be larger than the infinite-chain exponent. Since the same chain lengths were used, we suggest two possible explanations for these systematic differences. As we shall later show for these intermediate chain lengths, the weaker density-density correlation of finite smooth chain and finite open chain is not an artefact resulting from spatial averaging done to `restore' translational symmetry.  Instead, it is a genuine finite size effect related to the choice of smooth \cite{Munder2009NJP0910} and open boundary conditions, as opposed to periodic boundary conditions used on our finite periodic chains.

\subsection{Pair-pair correlations}
The pair-pair correlation function $\braket{\Axd{j-2}\Axd{j}\Ax{j+r}\Ax{j+r+2}}$
is also identical for both spinless fermions and hardcore bosons.  For spinless
fermions, $\braket{\Axd{j-2}\Axd{j}\Ax{j+r}\Ax{j+r+2}}$ can be interpreted as
the superconducting correlations.  As shown in Fig.~\ref{fig:scpo}, the pair-pair
correlation function can be fitted very well to the asymptotic form $A_1 r^{-A_2} + A_3 r^{-A_4} \cos (kr + \phi)$.  As expected, the wave number is $k = 2\pi\bar{N}$ if we allow $k$ to be a fitting parameter.  Again, from Fig.~\ref{fig:scpofp}, we see that there is reasonable agreement on the amplitudes and exponents between the finite and infinite chains, but poor agreement between their phase shifts.

\begin{figure}
\centering
\subfigure{\label{fig:sc1003}\includegraphics[width=0.45\textwidth]{scf3l}}
\subfigure{\label{fig:scol3}\includegraphics[width=0.45\textwidth]{scol3}}
\caption{Pair-pair correlation function
$\braket{\Axd{j-2}\Axd{j}\Ax{j+r}\Ax{j+r+2}}$ for particles with infinite
nearest-neighbor repulsion on (a) a periodic (b) an open chain of length $L = 100$ as a function
of the separation $r$. In this figure, red circles, blue squares, and black
triangles are the exact correlations at densities (a) $\Nbar = 0.20, 0.25, 0.30$ (b) $\Nbar=0.05, 0.10, 0.15$, respectively. The red, blue and black curves are the best nonlinear fits of the
form $A_1\, r^{-A_2} + A_3\, r^{-A_4}\, \cos(kr + \phi)$ to the exact correlations at these
densities.}
\label{fig:scpo}
\end{figure}

\begin{figure}
\centering
\includegraphics[scale=0.325,clip=true]{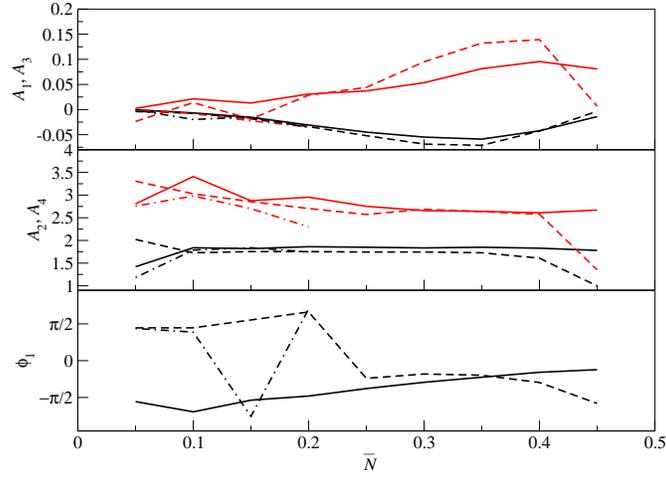}
\caption{The fitted amplitudes $A_1$ and $A_3$ (top), exponents $A_2$ and $A_4$ (middle), and phase shifts $\phi$ (bottom) of the pair-pair correlation function of a $L = 100$ periodic(dashed lines) and open(dot-dashed lines) chain of particles with infinite nearest-neighbor repulsion, as a function of the density $\Nbar$. The corresponding parameters of the infinite chain from Ref.~\cite{Cheong2009PRB80e165124} are shown as solid lines.}
\label{fig:scpofp}
\end{figure}

\subsection{Comparison between different chain lengths}
\begin{figure}[htbp]
\centering
\includegraphics[scale=0.3,clip=true]{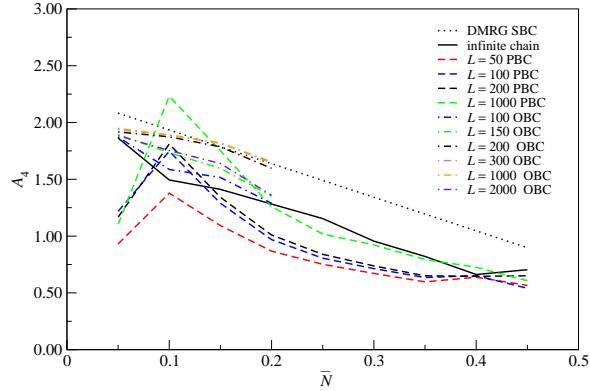}
\caption{The fitted oscillatory power-law exponent $A_4$ for the density-density
correlation function of hardcore particles with infinite nearest-neighbor
repulsion as a function of density $\Nbar$, for different periodic chain
lengths, $L = 50$ (red), $L = 100$ (blue), and $L = 200$ (black).  The dashed
line is the infinite-chain exponent from
Ref.~\cite{Cheong2009PRB80e165124}, whereas the dotted line is the
best-fit straight line to the finite-smooth-chain exponent from
Ref.~\cite{Munder2009NJP0910}.}
\label{fig:cdwfss}
\end{figure}

In the DMRG study of finite smooth chain \cite{Munder2009NJP0910}, a systematic study on the effects of
chain lengths was undertaken, even though there was no rigorous attempts at
performing finite size scaling.  They found, for the oscillatory power-law
exponent $A_4$ of the density-density correlation on ladders of lengths $L =
100, 150, 200$, the fitted parameter appears to have converged onto the
straight line $A_4 = -2.96\,\Nbar + 2.23$.  In Fig.~\ref{fig:cdwfss}, we show
the fitted exponent of the density-density correlations on finite periodic
chains with lengths $L = 50, 100, 200, 1000$ and finite open chains with lengths $L=100, 150, 200, 300, 1000, 2000$.  For the finite periodic chain, we also find an apparent convergence at a chain length of approximately $L = 200$. However, this `limiting' behaviour of $A_4$ does not agree with the exponent determined
from the infinite chain, and is also systematically different from the `limiting' finite-smooth-chain exponent.  In fact, unlike the infinite-chain and finite-smooth-chain exponents, the finite-periodic-chain exponent appears to be a strongly nonlinear function of $\bar{N}$. When we go to a $L = 1000$ periodic chain, the exponent $A_4$ is still strongly $\bar{N}$-dependent. However, we can see from Fig.~\ref{fig:cdwfss} that it is closer to the infinite-chain limit.

For finite open chains, however, the chain-length-dependence of the exponent $A_4$ is rather different.  Starting from being close to the infinite-chain limit at $L = 100$, we see the exponent moving \emph{away} from the infinite-chain limit as the chain length goes to $L = 150$ to $L = 200$ to $L = 300$ to $L = 1000$.  In fact, there is an apparent convergence at $L = 200$ to the finite-smooth-chain exponent.  This suggests that a $L = 100$ chain with smoothed open boundary conditions behaves like a longer chain with hard open boundary conditions, which is what they desired when White introduced the smoothed boundary conditions \cite{Vekic1993PRL71e4283, Vekic1996PRB53e14552}.  What they, and DMRG group would not have guessed, is the finite open chain exponent deviating from the infinite chain exponent initially as we go to longer and longer chains. From the open chain results of $L = 1000$ and $L = 2000$, we find that the convergence to infinite-chain results is slower than for finite periodic chains.

With this, we demonstrated that the choice of periodic/open boundary conditions resulted in the finite-chain exponents being systematically lower/higher than the infinite-chain exponent, for chain lengths on the order of $L = 100$ to $L = 200$.  This observation is important, because DMRG calculations on finite smooth chains are currently limited to such chain lengths. Furthermore, we showed that the finite-chain exponents do converge onto the infinite-chain limit, but only when the chain length is on the order of $L = 1000$. Moreover, this convergence is not uniform for finite chains. This poses an important challenge for the use of finite size scaling to determine the infinite-chain limits.

\begin{figure}[htbp]
\centering
\includegraphics[scale=0.3,clip=true]{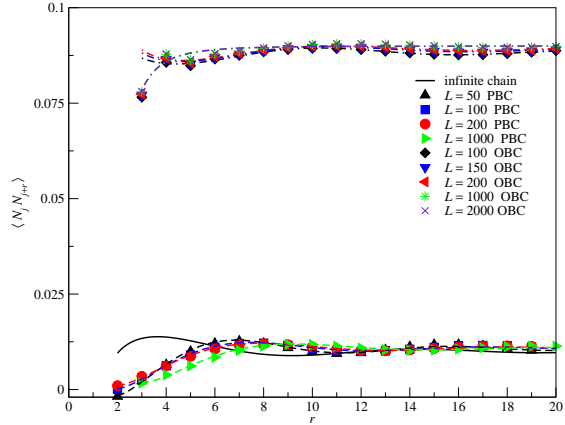}
\caption{The density-density correlation functions of infinite, finite periodic, finite open chains with $\Nbar = 0.10$. The solid line is the infinite-chain exponent from Ref.~\cite{Cheong2009PRB80e165124}.}
\label{fig:cdwfcfss}
\end{figure}

Finally, we compare the density-density correlation functions of different chain lengths, against the infinite-chain density-density correlation function, at density $\bar{N} = 0.10$.  As shown in Fig.~\ref{fig:cdwfcfss}, oscillatory power-law decay in the density-density correlation function is seen for finite and infinite chains.  The finite periodic chain and infinite chain amplitudes are very similiar.  The rates at which the density-density correlations decay on the finite periodic chain and infinite chain are also very similar.  This explains the reasonable agreement seen in the fitted amplitudes and exponents.

However, the phase shifts in the finite periodic chain and the infinite chain are rather different, and increasing the chain length from $L = 50$ to $L = 100$ to $L = 200$ does not make the finite-chain phase shift approach the infinite-chain phase shift.  This is true even at $L = 1000$.  As pointed out by the referee, this is expected as the phase shift is a parameter that is adjusted globally to minimize the effects of the boundaries, and thus will not converge smoothly with increasing chain length.

More importantly, the amplitude of the oscillatory power law is very small for finite open chains, because of the spatial averaging performed.  This same spatial averaging is done in the finite-smooth-chain calculations.

\section{Conclusions}
\label{sect:conclusions}
In conclusion, we employed the exact mapping between the ground states of the
excluded chains and included chains of hardcore bosons and spinless fermions,
along with the intervening-particle expansion method, both developed in
Ref.~\cite{Cheong2009PRB80e165124}, to investigate how the finite chain
length and the choice of the boundary conditions manifest themselves in the
two-point, density-density, and pairing correlation functions.  We fitted these
finite-chain correlations to the generic asymptotic form $A_0 + A_1\, r^{-A_2} +
A_3\, r^{-A_4}\, \cos(kr + \phi)$, and compared the fitted parameters to those
obtained from an infinite chain, and a finite smooth chain. In general, we find reasonable agreement between fitted amplitudes
and exponents for the finite and infinite chains, but poor agreement for the
fitted phases. We also find the phase shifts depend on both boundary conditions and system length.

Comparing the finite-periodic-chain, finite-open-chain, finite-smooth-chain and infinite-chain exponents for correlation functions at the same length,
we find that the finite-periodic-chain exponents are closer to the infinite-chain results and finite-open-chain exponents are closer to the finite-smooth-chain ones. We attribute this systematic difference to the finite size effects and the different choices of boundary conditions.  In a nutshell, finite-periodic-chain
correlations are stronger than they are with open or smooth boundary conditions (or any variants thereof).  More importantly, we observed at intermediate chain lengths an apparent `convergence' of our finite-chain exponents to functions of the density $\Nbar$ different from the infinite-chain exponents.  A
straightforward finite size scaling may thus lead us to the wrong physics.  This
same phenomenon was observed in the DMRG study of finite smooth chain as well, suggesting in general
that uniform convergence cannot be expected for chain lengths of $L \sim 10^2$, and perhaps not even for chain lengths of $L \sim 10^3$.
A more positive take on these general results would that that numerical studies
using more than one set of boundary conditions offer us a better sense of how
far we might be away from the infinite-system limits, even when there is no
guarantee that the results from different boundary conditions will bound the
true infinite-chain limit.

\section*{Acknowledgments}
This research is supported by startup grant SUG 19/07 from the Nanyang
Technological University. We thank Pinaki Sengupta for discussions, and the anonymous referee for suggestions on how to improve the paper.


\end{document}